\documentclass[prl, showpacs, superscriptaddress, twocolumn, floatfix, amsmath]{revtex4}
\usepackage{color, graphicx, amssymb}  

\date{07/28/09}
  
\begin{document}  
\title{Spin-like susceptibility of metallic and insulating 
thin films at low temperature}  
\author{Hendrik Bluhm}
\altaffiliation[Present address: ]
{Department of Physics, Harvard University, Cambridge, MA 02138, USA.}
\affiliation{Departments of Physics and Applied Physics, Stanford  
University, Stanford, CA 94305}  
\author{Julie A. Bert}  
\affiliation{Departments of Physics and Applied Physics, Stanford  
University, Stanford, CA 94305}  
\author{Nicholas C. Koshnick}  
\affiliation{Departments of Physics and Applied Physics, Stanford  
University, Stanford, CA 94305}  
\author{Martin E. Huber}  
\affiliation{Departments of Physics and Electrical Engineering,
University of Colorado Denver, Denver, CO 80217}  
\author{Kathryn A. Moler}  
\email{kmoler@stanford.edu}  
\affiliation{Departments of Physics and Applied Physics, Stanford  
University, Stanford, CA 94305}

\begin{abstract} 
Susceptibility measurements of patterned thin films at sub-K
temperatures were carried out using a scanning SQUID microscope that
can resolve signals corresponding to a few hundred Bohr magnetons.
Several metallic and insulating thin films, even oxide-free Au films, show a
paramagnetic response with a temperature dependence that indicates
unpaired spins as the origin. The observed response exhibits a
measurable out-of-phase component, which implies that these spins will
create $1/f$-like magnetic noise.  The measured spin density is
consistent with recent explanations of low frequency flux noise
in SQUIDs and superconducting qubits in terms of spin fluctuations,
and suggests that such unexpected spins may be even more ubiquitous
than already indicated by earlier measurements. 
Our measurements set several constraints on the nature of these spins.
\end{abstract}


\maketitle

The origin of 1/$f^\alpha$ noise in superconducting quantum
interference devices (SQUIDs) has been an unresolved mystery for more
than two decades. Part of this noise of typically a few
$\mu\Phi_0/\sqrt{\textrm{Hz}}$ at 1 Hz appears to be surprisingly
universal and behaves in every respect like true flux
noise \cite{WellstoodFC:LownDs,KochRH:Inv1nt}.
Measurements of electron spin dephasing
rates of donors in Si also give evidence for magnetic noise
originating at or near the surface \cite{SchenkelT:Eleaes}.
Recently, evidence that
a similar noise causes dephasing in superconducting qubits
\cite{YoshiharaF:Decfqd,BialczakR:Flunjp,HarrisR:Profnf,LantingT:Geodlf} 
has increased the interest in this
phenomenon. Koch, DiVincenzo and Clarke showed that fluctuating 
electron spins could explain the observed magnitude of
noise \cite{KochRH:Mod1fn}. Measurements of a flux offset in SQUIDs 
proportional to 1/$T$ \cite{SendelbachS:Magism} provide direct evidence for 
the presence of spins in superconducting devices. 
Even though they are likely related to material imperfections such as
surface oxides, defect states or contaminations, the nature of these
hypothetical spins is interesting for its own sake, and mitigating
their effects is essential for several solid-state approaches to
quantum computation. It is important to understand their origin
and polarization dynamics, which determine the magnetic noise spectrum.

Various models assuming different relaxation mechanisms of unpaired
electron spins on defects were recently proposed. 
Koch {\em et al.} argued that the spin of an electron in a charge trap
could remain locked until it leaves the trap \cite{KochRH:Mod1fn}.
A different model exploring thermally activated, nonmagnetic two level
systems as cause for spin flips was motivated by the argument that
only a small fraction of all defects have an activation energy low
enough to allow charge fluctuations \cite{SousaR:Dansrm}.
Faoro and Ioffe  explored noise
from spin diffusion mediated by RKKY coupling via the conduction electrons 
in metallic device elements \cite{FaoroL:Micolf}.

We have measured the magnetic response of metallic and insulating
thin films from $T$ = 25 mK to 0.6 K, using a SQUID susceptometer in a 
scanning microscope.  The most prominent result is a
surprisingly large paramagnetic susceptibility with a $1/T$-like
temperature dependence, and a magnitude consistent with a spin-1/2 density
of about $4 \cdot 10^{17}$/m$^2$, close to estimates from 
$1/f^\alpha$ noise levels in SQUIDs.
Furthermore, the response has a measurable out-of-phase component, 
which implies polarization noise through the 
fluctuation-dissipation theorem (FDT).
Our results thus demonstrate the existence of paramagnetic spins, 
with a density and dynamics suitable for producing $1/f^\alpha$ noise.
Similar susceptibilities seen for Ag films and Au films with 
and without a sticking layer, together with previous results inferring the
presence of spins in superconducting devices, suggest that the spins can occur 
similarly for different materials.

\begin{figure}
\includegraphics[width=8.6cm]{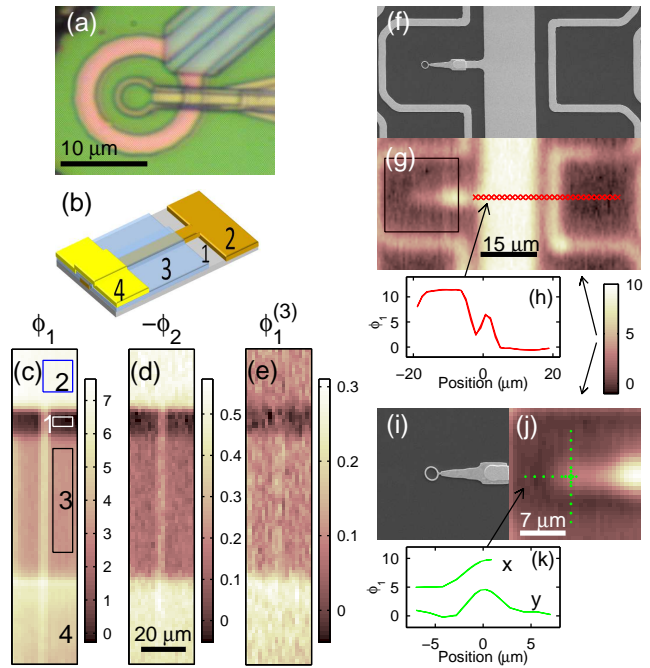}
\caption{\label{fig:scans}(color online)
(a) SQUID field coil and pickup loop.
(b) Schematic of the layer structure of sample II:
 1, Bare Si; 2, Au with Al adhesion layer; 3, ALD
deposited AlO$_x$; 4, Au on AlO$_x$.
(c)-(e) Sample II, linear in- and out-of phase
signal ($\phi_1$, -$\phi_2$) and in-phase 3rd harmonic ($\phi_1^{(3)}$)
at 193 Hz, 43 mK. All numbers in this figure are in units 
of $\mu\Phi_0$/mA and the response over bare Si has been defined as 0.
(f) Scanning electron micrograph of a region of sample I, 
Au films on Si with native oxide.  
(g) Sample I:  linear in-phase signal ($\phi_1$) at 193 Hz, 27 mK of a region
as shown in (f).
(h) Line scans over the positions indicated in (g), at 25 mK and 111 Hz. 
Panels (i)-(k) are the same as (f)-(h), zoomed in on ring as indicated by 
the box in (g).
The ring has a 2 $\mu$m diameter, 350 nm line width,
and a connection to the 15 $\mu$m wide wire for heat sinking.
The line scans in (k) were taken at 35 mK and the $x$-scan is offset for 
clarity.
The temperature dependence shown in Fig. \ref{fig:plots} (a) and (b) 
was obtained from line scans as in (h) and (k), or by averaging over the 
rectangles in (c).}
\end{figure}

We focus on results from two samples, which were designed
for other experiments \cite{Bluhm:Percnm} and include a range of structures
with different layer combinations. On sample I, rings and
wires were e-beam evaporated at a rate of about 1.2 nm/s from a 6N
purity Au source onto a Si substrate with a native oxide 
[Fig. \ref{fig:scans} (f), (i)].  First,
the wires, with widths of 2 and 15 $\mu$m, were patterned using optical
lithography and liftoff. Their thickness was 100 nm, including a 7 nm
Ti adhesion layer.  
Subsequently,  the micron-scale, 140 nm thick Au rings, which did not include
any adhesion layer, were defined using 
e-beam lithography with PMMA [poly(methyl methacrylate)] resist and liftoff.  
Some of them were connected to the wider wires for heat sinking.
Finally, Al rings for calibration purposes were fabricated in a similar way.
Before each metal deposition, the developed resist was descummed in 
an oxygen plasma.
The base pressure of our evaporator, which has never been used for magnetic 
materials, was below $5 \times 10^{-7}$ Torr. 
On sample II, the first layer, an e-beam defined, 80 nm thick
Au wire grid and  bonding pads, was evaporated onto a Si substrate 
with native oxide from a source with unknown purity on top of a 1 nm  
Al wetting layer.
A 50 nm thick AlO$_x$ film, patterned using optical lithography
and liftoff, was then deposited by atomic layer deposition (ALD).
Rings and heat sinks similar to those on sample I were fabricated
on top of the AlO$_x$ [Fig. \ref{fig:scans} (b)], also without adhesion
layer.

Our dilution-refrigerator based microscope \cite{BjornssonPG:Scasqi} 
employs SQUIDs \cite{HuberM:Susc} with an integrated 14 $\mu$m mean diameter 
field coil that is concentric with a 4.6 $\mu$m pickup loop 
[Fig. \ref{fig:scans} (a)]. These loops can be brought to within about 
1 $\mu m$ of the sample surface. The field coil applies an 
ac field $H_a$ (35 G amplitude at its axis for most of the data discussed 
here, corresponding to a field coil current $I_\textrm{FC}$ = 35 mA) 
to the sample, whose response couples a flux $\Phi_{SQUID}$ into the 
pickup loop.
A second, counter-wound pair of coils, located further from the 
sample, cancels the response of the SQUID to the applied field to within 
one part in $10^4$ \cite{HuberM:Susc}.
As the field coil current varies sinusoidally in time (with amplitude 
$I_\textrm{FC}$), the SQUID response $\Phi_\textrm{SQUID}$ 
is conveniently  characterized  in terms of its complex $n$th harmonics, 
$\Phi^{(n)}$.
We define $\phi_1^{(n)} + i \phi_2^{(n)} \equiv \Phi^{(n)}/I_\textrm{FC}$ and 
abbreviate $\phi_{1, 2} \equiv \phi^{(1)}_{1, 2}$. $\phi_1$ 
and $\phi_2$ quantify the in-phase and out-of-phase linear response,
$\phi_1^{(3)}$ is proportional to the cubic component.

\begin{figure}
\includegraphics{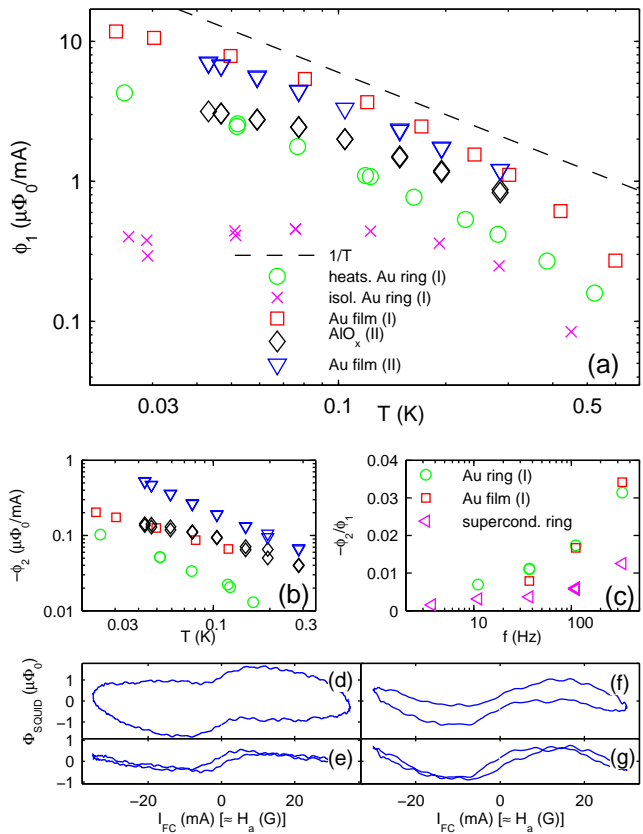}
\caption{\label{fig:plots}(color online)
(a) Temperature dependence of the in-phase linear susceptibilities ($\phi_1$)
 obtained from scans as shown in Fig. \ref{fig:scans} (see text) at 
111 Hz or 193 Hz.  Numbers in parenthesis indicate the sample.
The downturn at the highest $T$ can be attributed to 
the $T$ independent diamagnetic bulk susceptibility of 
$-3.4\cdot10^{-5}$ for Gold \cite{hbcp}.
(b) Out-of-phase component ($\phi_2$) from the same data sets.
(c) Frequency dependence of $\phi_2/\phi_1$ from
sample I at 25 mK. The data from the superconducting ring characterize 
the phase shift due to the finite measurement bandwidth and show that
the nonzero $\phi_2$ is not a measurement artifact. 
(d), (f) Nonlinear part of the response of the heatsunk Au ring (I), deposited 
directly on the Si substrate, at base temperature. (f), (g) Same for an 
isolated ring on sample II, deposited onto the AlO$_x$ film.
In (d), (f), only $\phi_1$ (i.e. a line) was subtracted. 
In (e), (g), both $\phi_1$ and $\phi_2$ (i.e. an ellipse) were subtracted. 
}
\end{figure}

Fig. \ref{fig:scans} shows 2D susceptibility scans of both samples.
For sample II, we took scans as shown in Fig. \ref{fig:scans}(c)-(e)
at a range of temperatures and extracted the temperature dependence
[Fig. \ref{fig:plots} (a), (b)] by averaging the indicated rectangular
regions.  For sample I,
we averaged the complete $I_\textrm{FC}$--$\Phi_\textrm{SQUID}$ curves
from many sinusoidal field sweeps at discrete positions as indicated
in Figs. \ref{fig:scans}(g), (j). 
This procedure is more sensitive and
allows us to determine the difference between the full responses
including nonlinearities [Fig. \ref{fig:plots}(d)-(g)] near and away
from the metal.
Figs.  \ref{fig:scans}(h), (k) show $\phi_1$
extracted from the response curves at each point. The $T$ and 
frequency dependencies of the magnitude of the spatial variation along these
line scans are included in Fig. \ref{fig:plots}(a)-(c).

While the above measurements can only detect lateral {\em variations}
of the sample response, the height dependence of the latter confirms
that the observed signals reflects a response from the metal film.
The height dependence over a third sample with a 1 $\mu$m thick
SiO$_x$ layer, grown at 1000$^\circ$C with a wet process, and above Si
with only a native oxide give evidence for a paramagnetic 1/$T$
surface response that is about a factor 5 and 30 smaller than than
that of the metal films, respectively  \cite{supp}. The response from
the ALD grown AlO$_x$ film, which likely has a higher defect density
than the thermal SiO$_x$, is comparable to that from the metal films
[Fig. \ref{fig:plots}(a)].  Comparison with the value of
$\phi_2/\phi_1$ for superconducting rings [Fig. \ref{fig:plots}(c)],
shows that the nonzero value of $\phi_2$ is not an instrumental
artifact.  Nevertheless, the actual sample contribution to $\phi_2$ is
somewhat smaller than the raw values displayed in
Fig. \ref{fig:plots}(b) \cite{supp}.

The $1/T$ dependence and the paramagnetic sign
of the susceptibility $\chi(\omega) \equiv \chi_1 + i\chi_2$ 
indicate that it originates from localized spins.
One can show that the $z$-field emanating from a film of thickness $d$ with 
an isotropic linear response is 
$\mathbf{B} = \mu_0 \chi  d \partial \tilde{\mathbf{H}}/\partial z$, 
where $\tilde{\mathbf{H}}$ is the  applied field reflected about 
the film ($xy$) plane. Using the measured
pickup-loop--field-coil inductance and modeling $\tilde{\mathbf{H}}$
as the field of a thin loop leads to 
$\chi_{1,2} d$ = 8 $\mu$m$\cdot$mA/$\Phi_0 \cdot \phi_{1,2}$.
For the films, this implies $\chi_1 T = 3\cdot10^{-5}$ K to within a
factor of 2. We estimate systematic errors of less than a factor 2
due to the simplicity of the model and uncertainties in the scan height.
The response of the ring is consistent with this estimate within its somewhat 
larger calibration uncertainty.
Comparing  $\chi_1 T = 3\cdot10^{-5}$ K to the susceptibility
$\chi_1 = \mu_0 n (g \mu_B)^2  J(J+1)/3 k_B T$ of dilute spins with number 
density $n$ and total angular momentum $J$ leads to a concentration 
of 60 ppm for $d$ = 100 nm, $g = 2$ and $J = 1/2$, corresponding to an 
area density of $4 \cdot 10^{17}$ spins/m$^2$. Because all our films
have a similar thickness, we cannot distinguish whether the 
spin density scales with the volume or surface area.
For common magnetic ions with $g^2 J(J+1) \approx 35$ \cite{AshcroftN}, 
the concentration would still be about 3 ppm, 
which is an order of magnitude larger than the specifications of 
our source material. For the materials investigated, the calculated 
equilibrium response from nuclear spins is several orders of magnitude 
smaller than the observed signals. 

The magnetic moment noise spectral density from a sample of volume $V$
and susceptibility $\chi$ that couples to a sensing loop as a dipole
is $S_m(\omega) = - 2 k_B T \chi_2( \omega) V/\pi \omega$.  This form
of the FDT was verified for a spin-glass, using a SQUID susceptometer
similar to ours \cite{ReimW:Magens}. In our case, the expected
magnetic noise from the sample is much too small to be detected
directly.  However, Ref. \onlinecite{KochRH:Mod1fn} shows that a spin
density similar to our estimates can explain the observed $1/f^\alpha$
noise levels in SQUIDs.  Integrating the dynamic susceptibility of
fluctuating spins over the relaxation time distribution assumed in
Ref. \onlinecite{KochRH:Mod1fn} leads to a value of $\phi_2/\phi_1$
that is also of the same order as our results.  Although we cannot
prove that $\phi_1$ and $\phi_2$ are of the same origin, the similar
$T$ dependence, and consistency of the signs with a lag due to a
finite relaxation rate, do suggest a direct connection. While it is
also not a priori clear to what extent our results apply to other
metals including superconductors, the similar phenomenology of $1/f$
flux noise, our data from Au and Ag films, and recent measurements on
superconducting devices also showing a $1/T$ susceptibility component
\cite{SendelbachS:Magism} indicate that all these effects are closely
related.

The frequency dependence of $\chi_2$ and the nonlinearities imply a
millisecond-scale spin relaxation time for some 
spins, which indicates weak coupling to the conduction electrons. 
On the other hand, we find that the linear susceptibility 
of isolated metal rings saturates below approximately 150 mK 
[Fig. \ref{fig:plots} (a)].
In such rings, the electrons are expected not to cool below that temperature
because of heating by Josephson oscillations in the 
SQUID \cite{Bluhm:Percnm}. This observation indicates that the spins 
thermalize with the  electrons rather than the lattice, which suggests
an electronic relaxation mechanism 
\footnote{This argument relies on the 
assumption that the interface between the ring and the substrate 
is of sufficient quality to allow efficient phonon transmission.}.

One may thus suspect a connection with evidence for  
spin impurities in metallic nanostructures and at surfaces and interfaces
obtained from transport measurements. Enhancement of superconductivity 
in nanowires in an applied field indicates pair
breaking by spins \cite{RogachevA:Magesu,WeiTC:Enhsmi}. Weak
localization measurements, which are a very sensitive probe for
magnetic impurities \cite{PierreF:Depemm}, show that TiO$_x$ adhesion
layers for Au wires \cite{TrionfiA:StrmsT} and native oxides on Cu
films \cite{VrankenJ:Enhmss} can cause spin-flip scattering.  
In contrast, our observation of similar susceptibilities from Au films 
with (wires sample I) and  without Ti layer (rings I and II) 
shows that TiO$_x$ is not the dominant source of spins in our samples. 
From standard 
weak localization measurements for $T\ge$ 300 mK on wires fabricated 
together with the samples discussed above, we find a dephasing rate
$1/\tau_\phi$ with a temperature dependence close to $1/\tau_\phi \propto
T^{2/3}$, as expected for electron-electron interaction mediated 
dephasing \cite{AleinerI:Intepr}.   The deviation from this power law behavior 
can be accounted for with 0.1 ppm of Mn impurities, if one allows
the prefactor of the $T^{2/3}$ term to be a factor 6 larger than
theoretically expected. In typical weak localization measurements, the 
discrepancy between the theoretical and experimental prefactor is no larger
than a factor two \cite{PierreF:Depemm}.
However, if the unusually large discrepancy in our case were due to spins, 
their Kondo temperature would have to be larger than about 1 K in order 
to explain the observed increase of $\tau_\phi$ at low $T$. 
While we cannot rule out the existence of such spins, they cannot explain the 
susceptibility signal because their response would be quenched by the Kondo
effect at low $T$.
On the other hand, the measured $\tau_\phi \ge 1$ ns for $T \le$ 1 K 
sets a low upper bound on the 
spin flip scattering rate and thus exchange coupling and Kondo 
temperature of the spins contributing to the susceptibility response
\footnote{This also implies that the flattening of $\chi_1(T)$ at 
the lowest $T$ in Fig. \ref{fig:plots} for Au film I and Au ring I 
cannot be explained as Kondo-quenching of the magnetic moments.}. 
This upper bound does not necessarily rule out the RKKY coupling proposed 
in Ref. \cite{FaoroL:Micolf}, where it was suggested that the 
magnetic noise is due to RKKY-mediated spin diffusion. 
However, spin diffusion mediated by an isotropic interaction as considered 
in Ref. \cite{FaoroL:Micolf} conserves total angular momentum and thus the
total magnetic moment (assuming no $g$-factor variations).  Thus, the
observation of a paramagnetic response from isolated rings, 
which are smaller than the pickup-loop \cite{Bluhm:Percnm} and thus couple 
to our sensor mostly through their total magnetic moment, is inconsistent
with this diffusion model. Nevertheless, anisotropic spin-spin interactions, 
such as dipolar coupling, could in principle determine the relaxation 
dynamics.

We finally discuss a few anecdotal observations.
We observed
comparable values of $\phi_1$ on two other samples, one similar to
sample I, but on a Si substrate with an approximately 1 $\mu$m thick
wet thermal oxide \cite{supp}, and one similar to sample II, but with Ag
substituted for the top two Au layers. The Ag films showed a substantial
spatial variation of the magnetic response. This inhomogeneity 
could be due to an inhomogeneous surface oxidation or other chemical 
contamination.  We find that the contributions
from different layers on sample II are in general not additive, which
most likely means that the spin population is concentrated at surfaces
or interfaces.
The nonlinear response and $\phi_2/\phi_1$ vary significantly between
different samples and different layers [Fig. \ref{fig:plots}(d)-(g)].
Even though the nonlinearity seen in sample II is predominantly 
cubic, its  magnitude and $T$ dependence  
are inconsistent with the saturation of the equilibrium response at 
finite field.
The relatively small ratios of the hysteretic and nonlinear components
to $\phi_1$ indicates that the majority of the spins contributing to the 
latter relax fast compared to the measurement frequency. 
The increase
of $\phi_2$ with frequency is qualitatively consistent with $S_\Phi(\omega)$
varying slower than $1/\omega$, as observed in some
SQUIDs \cite{HuberM:Susc,WellstoodFC:LownDs}.

In summary, we have measured the susceptibility of micropatterned thin
films.  Different samples showed similar linear responses
corresponding to an area density of unpaired spins on the order of
$0.4$ spins/nm$^2$.  The spins on our Au films appear to be weakly
coupled to conduction electrons.  The out-of-phase component of the
susceptibility gives direct experimental evidence for the hypothesis
that the $1/f$ flux noise seen in SQUIDs and superconducting qubits is
due to fluctuating spins \cite{KochRH:Mod1fn}.

\acknowledgments{ This work was supported by NSF Grants No.
DMR-0507931, DMR-0216470, ECS-0210877 and PHY-0425897, and
the Center for Probing the Nanoscale (CPN), an NSF NSEC,
NSF Grant No. PHY-0425897. Work was performed in part at the Stanford 
Nanofabrication Facility, which is supported by NSF Grant No. ECS-9731293, 
its lab members, and industrial affiliates.}

\bibliography{susc_bibdata,pc_bibdata}
\end{document}